\documentclass[%
reprint, superscriptaddress, onecolumn,  aps,
 amsmath,amssymb,
]{revtex4-2}

\usepackage{graphicx}
\usepackage{dcolumn}
\usepackage{bm}

\newcommand{\beq}{ \begin{equation} }
\newcommand{\eeq}{ \end{equation} }
\newcommand{\beqs}{ \begin{eqnarray} }
\newcommand{\eeqs}{ \end{eqnarray} }

\usepackage{xcolor}

\usepackage[utf8]{inputenc}
\usepackage[T1]{fontenc}
\usepackage{mathptmx}

\begin{document}

\title{Pinch-off dynamics to describe animal lapping timescale} 

\author{Sunghwan Jung} 
\email{sunnyjsh@cornell.edu}
 \affiliation{Biological and Environmental Engineering, Cornell University, NY 14853, USA}

\date{\today}


\begin{abstract}
Some carnivorous mammals (e.g., cats and dogs) lap water with their tongues to drink water at high frequencies. Such a fast moving tongue creates a liquid column out of a bath which is bitten by the mouth for drinking. Presumably, the animals bite just before the pinch-off time of the water column to maximize the water intake. Otherwise, the water column falls back to the bath before being bitten. Such a pinch-off phenomenon in the liquid column can be described as the acceleration-induced (i.e., unsteady) inertia balances with the capillary force. The classical Rayleigh-Plateau instability explains the competition of the steady inertia with the capillarity, but not with the unsteady inertia. In this study, we modify the Rayleigh-Plateau instability in the presence of the fluid acceleration, and show that the most unstable wavenumber and growth rate increase with acceleration. The pinch-off time is theoretically predicted as the -1/3 power of the Bond number (i.e, a ratio of the acceleration-induced inertia to capillarity). Finally, measured pinch-off times from previous physical experiments and dog \& cat jaw-closing times are shown to be in good agreement with our theoretical pinch-off time. Therefore, our study shows that animals presumably modulate their lapping and jaw-closing time to bite down the water column before the pinch-off to maximize the water intake. 
\end{abstract}

\keywords{??}

\maketitle 

\section{Introduction}
Mammals are composed of 50--70\% water in the body \cite{RICHMOND1962,Calder1981}. The water content needs to be internally circulated, discharged by urination \cite{Yang2014}, and externally supplied by drinking water \cite{Reis2010,Gart2015}. Most mammals drink water using a suction mechanism. For example, humans drink or suck water by lowering pressure in the mouth, which is possible by sealing its mouth from the atmosphere. However, most carnivorous mammals cannot lower the pressure in the mouth for drinking due to the incomplete cheek. Instead, these animals develop a lapping mechanism as the tongue moves in and out of the water. While the tongue is pulled up, a water column is created due to the high inertia force and wettable tongue surface. As the water column is formed between the tongue and the free surface, it becomes unstable to break into two or several pieces (i.e., pinch-off). Before the pinch-off, animals need to bite down the portion of the column for drinking. By doing so, the animals can drink some amount of water even though the remaining water column falls back to the bath. Presumably, animals regulate the lapping frequency and biting time to maximize the water intake from a physics point of view. The pinch-off dynamics of a liquid column has been observed not only when cats and dogs lap water \cite{Reis2010,Gart2015} but also when aquatic animals jump out of water \cite{Kim2015a,Chang2019}. 

The Rayleigh-Plateau instability is one of the canonical examples in hydrodynamic stability \cite{Rayleigh:1879vw,Plateau1873}, which describes how a liquid column breaks into a series of small droplets (e.g., a water stream from a faucet). The pinch-off dynamics has been extensively studied in the cases of a liquid bridge \cite{Yildirim2001,CHEN:1997kq}, a flowing jet \cite{Rayleigh:1879vw,Utada2008,Sirignano2000} , and a water-exiting object \cite{Kim2015a,Chang2019}. We summarize the theoretical pinch-off time into three different cases. 
At extremely low speeds (i.e., quasi-static regimes), the pinch-off time can be predicted when the length of the liquid column becomes its capillary length (i.e., $\sqrt{\gamma/\rho g}$ where $\gamma$ is the surface tension, $\rho$ is the fluid density, and $g$ is the gravitational constant) \cite{Marmottant:2004dr}. The capillary length indicates the maximum length of the liquid column under gravity, which is typically on the order of a few milimeters. 
Next, at a constant speed of the column separation, the liquid column pinches off at the time of $(V/L)(V/\sqrt{gL})^{2/3}$ where $V$ is the separation or jet speed and $L$ is the characteristic length (i.e., typically a column diameter) \cite{Hao2020}. The term inside the 2/3 power is called ``Froude number'' as a ratio of inertia to gravitational force. This shows that the pinch-off time is determined by the time when the steady inertia balances with gravity.
Lastly, when the liquid column is stretched or flowing with acceleration, the pinch-off time is predicted as $\sqrt{L/a}$ where $a$ is the acceleration of a liquid column \cite{Vincent:2014il,Gart2015,Kim2018,Weickgenannt:2015kn}. To derive this theoretical pinch-off time, one of the main assumptions is a constant wavenumber determined by the column size, which has not been confirmed yet. 


In this present study, we will investigate the pinch-off time of a liquid column with acceleration. The acceleration-induced Rayleigh-Plateau instability is rationalized using the governing equations for 1D or 2D liquid columns. By linearizing the governing equations with proper boundary conditions, dispersion relations are obtained to find the most unstable mode. Then, the most unstable growth rate and wavenumber are used to predict the pinch-off time without using the assumption of constant wavenumber that is previously used. Finally, we compare our theoretical pinch-off time with previously reported experimental data (i.e., a sphere or a cylinder moving out of a liquid bath, dog and cat lapping videos).

\begin{figure*}
\centering
\includegraphics[width=1\textwidth]{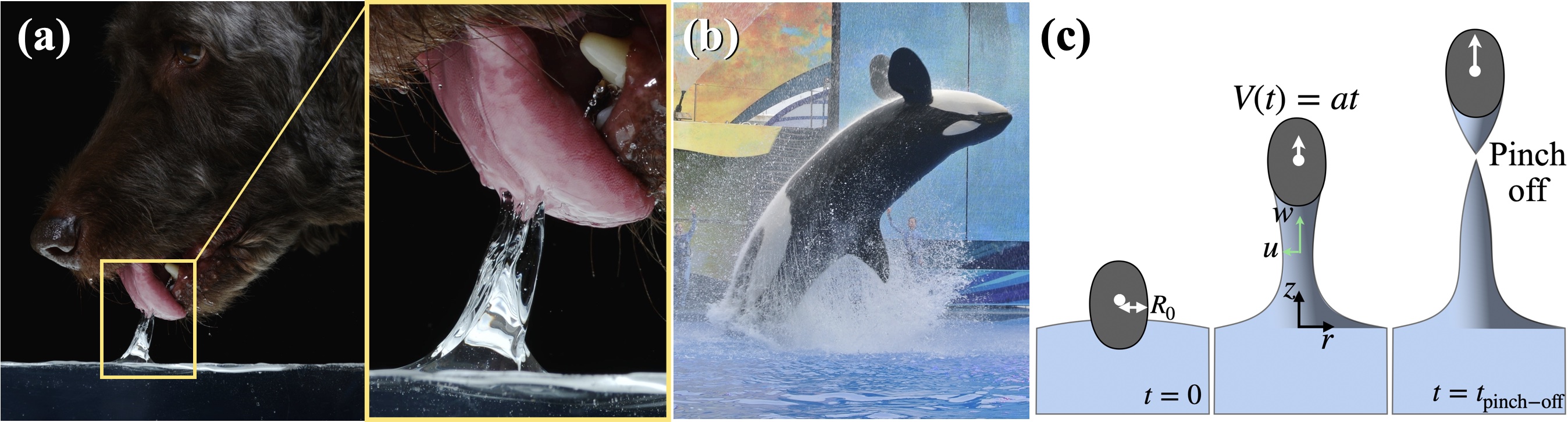}
\caption{(a) Photos show the formation of a water column while the dog drinks (Image credit to Jake Jocha, Sean Gart, and Sunghwan Jung). (b) Water columns are formed while the killer whale jumps out of water (Image credit to Sunghwan Jung). (c) Schematic of a liquid column underneath an object exiting the liquid surface. 
}
\label{Motiv}
\end{figure*}

\section{Results}
We present two different methods to solve for a dispersion relation and predict the pinch-off time when an object creates a liquid column out of a bath with acceleration. Similar calibrations without acceleration can be found in fluid mechanics books such as Chap 1.5 in \cite{drazin2004hydrodynamic}. Such a water-exit phenomenon with acceleration is similar to a water column in animal lapping and jumping behaviors as shown in Fig. \ref{Motiv}(a,b). Especially, the pinch-off phenomenon would play an important role in animal's lapping as a liquid column becomes unstable due to inertia and surface tension.

\subsection{2D Column \label{Sec:2d}}
When an axisymmetric water column is formed above the free surface by a water-exiting object at a high speed (see Fig. \ref{Motiv}(c)), we presume (1) negligible viscous effect and (2) no azimuthal dependence. Then, the Euler equations in cylindrical coordinates can be written as
\beqs 
\rho \left( \frac{\partial u}{\partial t} +u \frac{\partial u}{\partial r} + w \frac{\partial u}{\partial z}\right) &=& - \frac{\partial p}{\partial r} \\
\rho \left( \frac{\partial w}{\partial t} +u \frac{\partial w}{\partial r} + w \frac{\partial w}{\partial z}\right) &=& - \frac{\partial p}{\partial z} - \rho g \\
\frac{1}{r}\frac{\partial (r u)}{\partial r} + \frac{\partial w}{\partial z}&=&0 \,,
\eeqs
where $\rho$ is the fluid density, $u$ is the radial velocity of the fluid, $w$ is the axial velocity, $p$ is the pressure, and $g$ is the gravitational constant.  
We expand the variables as a base state and a perturbed state as
\beqs 
& u = 0+\delta u(r,z,t), ~~ &  w = at + \delta w(r,z,t), \nonumber \\ & R = R_0 - \delta R(z,t), ~~  &  p = p_\mathrm{atm}-\rho gz -\gamma/R_0^2 + \delta p(r,z,t) \,,
\eeqs 
where $a$ is the acceleration of the object, $R$ is the column radius, $R_0$ is the initial column radius, $p_\mathrm{atm}$ is the atmospheric pressure, and $\gamma$ is the surface tension. The base state of the radial velocity is assumed to be zero, whereas the base state of the axial velocity is to be ``$a\,t$'' as the body is moving out of water at constant acceleration, $a$.
Then, the linearized Euler equations become  
\beqs 
\rho \left( \frac{\partial }{\partial t} + at \frac{\partial }{\partial z}\right)\delta u =- \frac{\partial }{\partial r} \delta p \label{eq:2drdt}\\
\rho \left( \frac{\partial }{\partial t}  + at \frac{\partial }{\partial z}\right) \delta w =- \frac{\partial }{\partial z} \delta p\label{eq:2dzdt} \\
\left( \frac{\partial }{\partial r} + \frac{1}{r} \right) \delta u + \frac{\partial }{\partial z}\delta w=0 \,. \label{eq:cont2}
\eeqs
First, we take a $z$-derivative on Eq. (\ref{eq:2drdt}) and a $r$-derivative on Eq. (\ref{eq:2dzdt}). Then, we subtract one equation from the other to get rid of the $\delta p$-dependence as
\beqs 
&&\left( \frac{\partial }{\partial t} + at \frac{\partial }{\partial z}\right)\frac{\partial }{\partial z} \delta u = \left( \frac{\partial }{\partial t}  + at \frac{\partial }{\partial z}\right) \frac{\partial }{\partial r} \delta w \\
\rightarrow ~~&& \frac{\partial}{\partial z} \delta u = \frac{\partial }{\partial r} \delta w
\eeqs 
This linearized relation between $\delta u$ and $\delta w$ is used in Eq. (\ref{eq:cont2}) after taking a $r$-derivative. Then, the continuity equation becomes  
\beq
\left[ \frac{\partial }{\partial r} \left( \frac{\partial }{\partial r} + \frac{1}{r} \right)    + \frac{\partial^2 }{\partial z^2} \right] \delta u=0 \,.
\eeq 
This equation has solutions, the so-called modified Bessel functions of the first or second kind. To have a finite value at $r=0$, only the first kind is valid as $\delta u(r,z,t)=C\, I_1(kr)\exp(ikz - i\omega t)$ where $C$ is an unknown constant, $I_1$ is the modified Bessel function of the first kind, $k$ is the wavenumber, and $\omega$ is the frequency. The imaginary part of the frequency, $\mathrm{Im}[\omega]$, is also called ``growth rate'', which corresponds to the growth rate of variables over time. Similarly, the imaginary part of $k$ affects the magnitude of variables along $z$ as a spatial growth rate. 

To further solve these equations, two boundary conditions on the column surface (@ $r=R$) are used; 

\textbf{(1)} The first one is the kinematic boundary condition on the radial velocity as $u|_{r=R} = Dr/Dt|_{r=R} = \partial R /\partial t + w \partial R /\partial z$ where $D/Dt$ is the material derivative. The first order of the kinematic boundary condition reduces to 
\beq 
\delta u|_{r =R} = \left( \frac{\partial}{\partial t} + at \frac{\partial}{\partial z} \right) \delta R\,.
\eeq 

\textbf{(2)} The second boundary condition is the Young-Laplace equation as $p - p_\mathrm{atm} = \gamma \nabla \cdot \hat{n} = \gamma (1/R_1 + 1/R_2)$ where $\gamma$ is the surface tension, and $R_1$ \& $R_2$ are the radii of curvatures. The first curvature can be chosen to be the inverse of its own column radius ($1/R_1 \equiv 1/R = 1/(R_0-\delta R)=1/R_0 + \delta R/R_0^2 +{\cal O}(\delta R^2)$). Then, the second curvature can be chosen to be orthogonal to the first one, which becomes $1/R_2 \equiv ({\partial^2 R}/{\partial z^2}) /[1+|{\partial R}/{\partial z}|^{2}]^{3/2}$. Under the small slope assumption (i.e., $|{\partial R}/{\partial z}|\ll 1$), the second curvature can be approximated as ${\partial^2 R}/{\partial z^2}$. Finally, the first-order Young-Laplace equation becomes 
\beq 
\delta p = \gamma \left( \frac{1}{R_0^2} + \frac{\partial^2 }{\partial z^2} \right) \delta R \,.
\eeq 

Employing the above two boundary conditions and taking a derivative of $(\partial /\partial t + at \partial /\partial z)$ on Eq. (\ref{eq:2drdt}), one gets  
\beq
\rho \left( \frac{\partial }{\partial t} + at \frac{\partial }{\partial z}\right)^2 \delta u|_{r=R} =- \gamma \frac{\partial}{\partial r}  \left( \frac{1}{R_0^2} + \frac{\partial^2 }{\partial z^2} \right) \delta u|_{r=R} \,. \label{eq:2ddt}
\eeq 

Then, with the normal mode assumption ($\delta u|_{r=R}=CI_1(kR)\exp(ikz - i\omega t)$), the dispersion relation is obtained as  
\beq 
(\omega - atk)^2  = iak -\frac{\gamma}{\rho R_0^3} (kR_0) (1-(kR_0)^2) \frac{I_1(kR_0)}{I_0(kR_0)} \,.
\eeq 
Here, the first term, $iak$, on the right-hand side is due to the fact that ``$a\,t$'' in front of $\partial/\partial z$ on the left-hand side of Eq. (\ref{eq:2ddt}). When you take the material derivative twice in Eq. (\ref{eq:2ddt}), this term appears [see $( {\partial }/{\partial t }  + at {\partial }/{\partial z})^2 = {\partial^2 }/{\partial t^2 } + a {\partial }/{\partial z} + 2at {\partial^2 }/{\partial z \partial t} + (at)^2 {\partial^2 }/{\partial z^2}$]. 
The modified Bessel functions on the last term can be approximated as $I_\alpha(z) \sim (\Gamma(\alpha+1))^{-1} (z/2)^\alpha$. With this approximation (i.e., $I_1(kR_0)/I_0(kR_0) \sim kR_0/2$), the above dispersion relation is further simplified to an equation, which is the same as in Eq. (\ref{eq:1D_RP}) for a 1D column in the next section. 

\begin{figure}
\centering
\includegraphics[width=.9\textwidth]{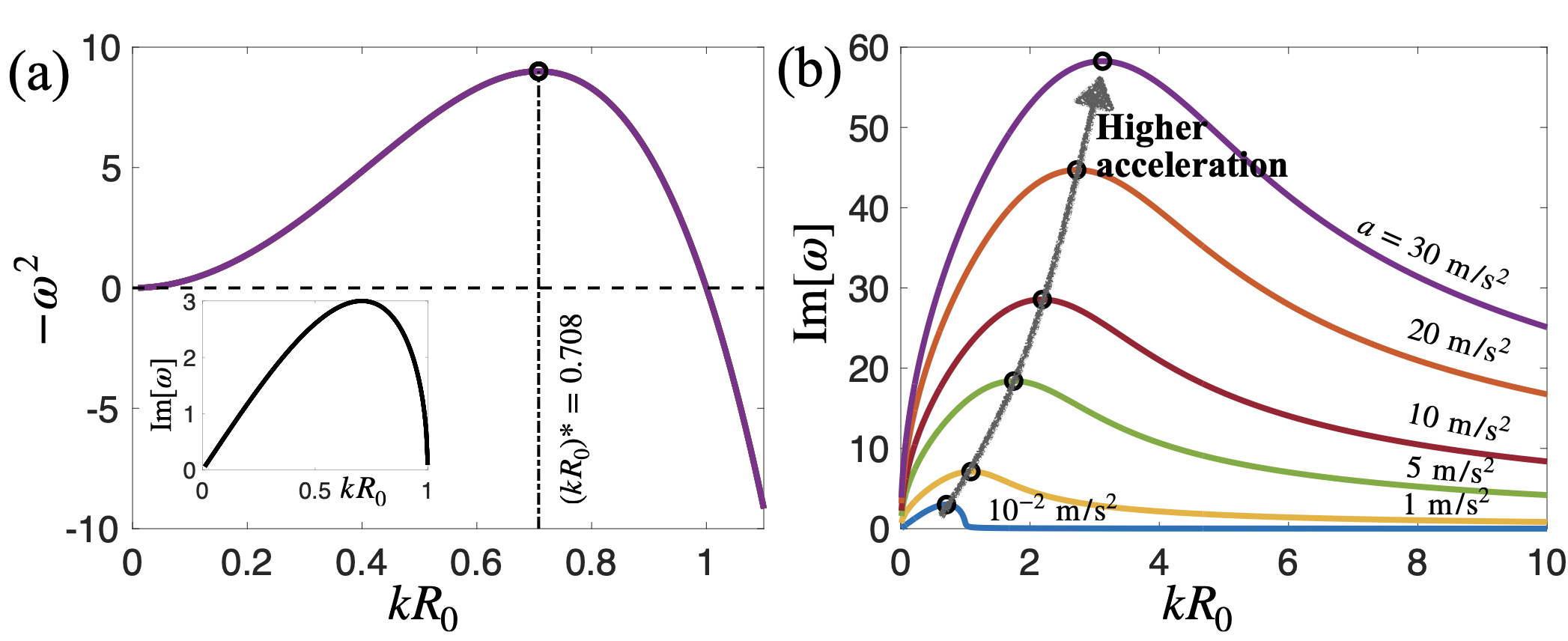}
\caption{(a) Dispersion relation of the classical Rayleigh-Plateau instability with $\gamma = 0.072$ N/m, $\rho = 1000$ kg/m$^3$, and $R_0 = 10^{-2}$ m. The growth rate is maximized at $(kR_0)^* = 0.708$ regardless of fluid properties. (b) Dispersion relation of the modified Rayleigh-Plateau instability with acceleration. Here, both the growth rate and non-dimensional wavenumber of the most unstable mode increase with acceleration. 
}
\label{Dispersion}
\end{figure}

\subsection{1D column}
We can simplify the Navier Stokes equations of a liquid column into 1D equations of the vertical velocity, $w(t,z)$, and the column radius, $R(t,z)$ only. This equation has been widely used in previous studies \cite{Nagel1994,Eggers1994,Eggers:2008hq,Kim2018}. The governing equations are  
\begin{eqnarray}
&& \rho \left( \frac{\partial w}{\partial t} +w \frac{\partial w}{\partial z} \right) = - \frac{\partial p}{\partial z} - \rho g\\ && \frac{\partial R}{\partial t} = -w \frac{\partial R}{\partial z} - \frac{R}{2} \frac{\partial w}{\partial z},
\end{eqnarray}
where $\rho$ is the fluid density, $p$ is the pressure, $g$ is a gravitational constant, and $R$ is the column radius. To linearize the above equation, we decompose the axial velocity, radius, and column pressure into a zeroth-order base state and a first-order perturbation as
\beq 
w = at + \delta w(z,t), ~~ R = R_0 - \delta R(z,t), ~~ p = p_0 + \delta p(z,t) \,.
\eeq 
Similar to the pressure calculation in \ref{Sec:2d}, the first-order pressure term from the Young-Laplace equation becomes
\beq
\delta p = \gamma \left( \frac{1}{R_0^2} + \frac{\partial^2}{\partial z^2} \right) \delta R \,.
\eeq
Then, the first order of the governing equations becomes
\begin{eqnarray}
{\rho} \left( \frac{\partial }{\partial t}  + at \frac{\partial }{\partial z} \right) \delta w &=& - {\gamma} \left( \frac{1}{R_0^2} \frac{\partial }{\partial z} + \frac{\partial^3 }{\partial z^3} \right) \delta R \label{eq:dt} \\ 
\left( \frac{\partial }{\partial t} + at \frac{\partial }{\partial z} \right)\delta R  &=&  \frac{R_0}{2} \frac{\partial }{\partial z} \delta w \label{eq:drdt} \, .
\end{eqnarray}
By taking a $z$-derivative on Eq. (\ref{eq:dt}), one gets
\beq
{\rho} \left( \frac{\partial }{\partial t }  + at \frac{\partial }{\partial z}  \right) \frac{\partial }{\partial z} \delta w = -{\gamma} \left( \frac{1}{R_0^2} \frac{\partial^2 }{\partial z^2} + \frac{\partial^4 }{\partial z^4} \right) \delta R \,.
\eeq 
Then, it is plugged into Eq. (\ref{eq:drdt}) as 
\beq
\left( \frac{\partial }{\partial t }  + at \frac{\partial }{\partial z}  \right)^2 \delta R = - \frac{R_0 \gamma}{2\rho} \left( \frac{1}{R_0^2} \frac{\partial^2 }{\partial z^2} + \frac{\partial^4 }{\partial z^4} \right) \delta R \,.
\eeq 
It is worth noting that the square of the material derivative on the left-hand side should be performed carefully as mentioned in \ref{Sec:2d}. Assuming the normal mode $\delta R = C \exp(ikz -i\omega t)$ and multiplying $R_0^4$, one gets the dispersion relation as 
\beq
 \left( \omega  -  at k  \right)^2  = iak - \frac{ \gamma}{2\rho R_0^3} \left[ (kR_0)^2  - (kR_0)^4  \right] \label{eq:1D_RP} \,.
\eeq 
In the limit of small acceleration as $a \rightarrow 0$, we recover the dispersion relation for the classical Rayleigh-Plateau instability.

\subsection{Pinch-off time}
In this section, let us consider the dispersion relation of the 1D column instead of the 2D column. The main reason is that we can obtain a simple analytical solution of the pinch-off time without Bessel functions. Also, we need one more simplification to calculate the pinch-off time; the convective frequency, $\omega - atk$, on the left-hand side of the dispersion relation represents the growth rate of a mode in the convected or moving frame while the fluid is stretched \cite{Keller1973,Eggers:2008hq}. To obtain the theoretical pinch-off time from the growth rate and the wavenumber, we do not have to consider the convective frequency as it is. Instead, we will use $\omega$ in lieu of $\omega -atk$ for convenience from now. 
Hence, the dispersion relation of Eq. (\ref{eq:1D_RP}) can be rewritten as   
\beq 
\omega^2 = i\frac{a}{R_0}kR_0 - \frac{ \gamma}{2\rho R_0^3} \left[ (kR_0)^2  - (kR_0)^4  \right] \,. \label{pinoff-disp}
\eeq 
On the right-hand side, the first term is the effect of acceleration and the second term shows the classical Rayleigh-Plateau instability. The classical Rayleigh-Plateau instability predicts that the most unstable mode is constant as $(kR_0)^* \simeq 0.708$ (see Fig. \ref{Dispersion}(a)). However, when we consider the effect of acceleration (i.e., the first term), numerical computation shows that the most unstable mode increases with acceleration (see Fig. \ref{Dispersion}(b)). 

\begin{figure}
\centering
\includegraphics[width=.9\textwidth]{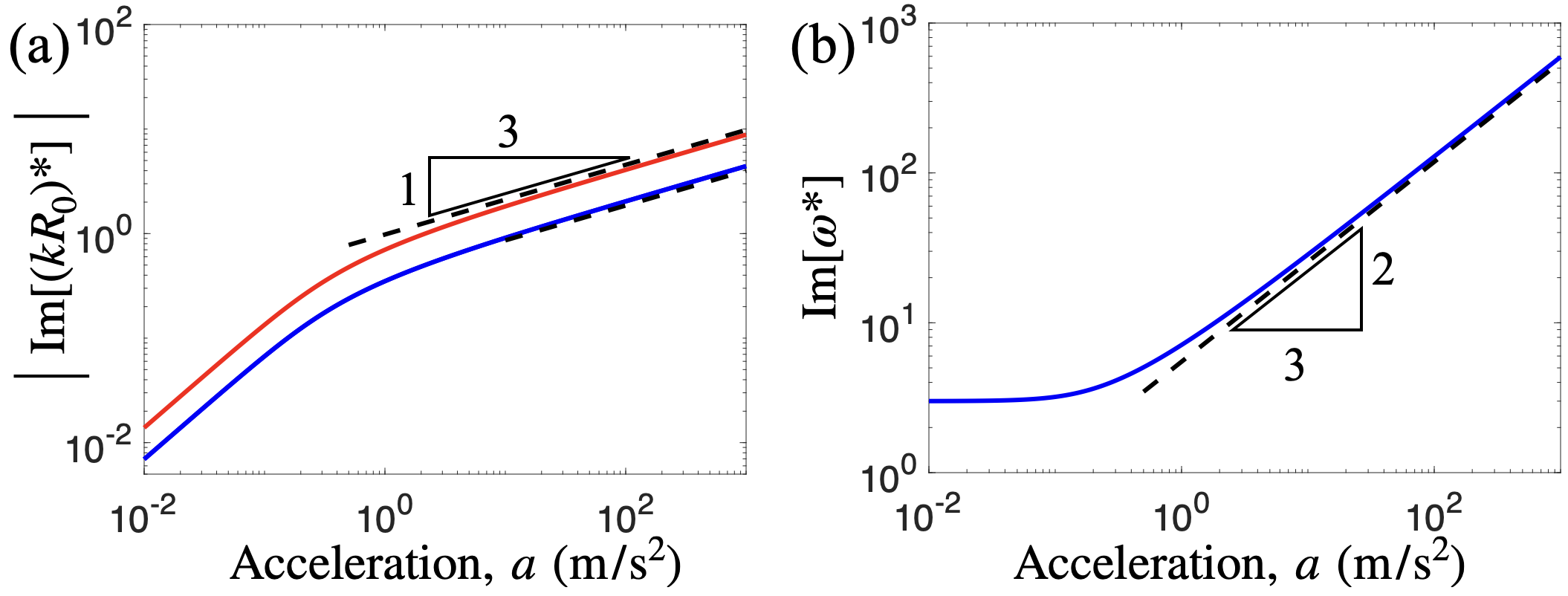}
\caption{ Fluid properties are $\gamma = 0.072$ N/m, $\rho = 1000$ kg/m$^3$, and $R_0 = 10^{-2}$ m. (a) Imaginary part of the most unstable non-dimensional wavenumbers ($\mathrm{Im}[(kR_0)^*]$) scales as the 1/3 power of acceleration: $\mathrm{Im}[(kR_0)^*] \propto a^{1/3}$. Three solutions are obtained from Eq. (\ref{eq:maxima}); One red line and two blue lines. (b) The most unstable growth rate ($\mathrm{Im}[\omega^*]$) scales as the 2/3 power of acceleration: $\mathrm{Im}[(\omega)^*] \propto a^{2/3}$. 
}
\label{most}
\end{figure}

The local maxima (i.e., the most unstable mode) occurs when 
\beqs 
&& 0 = \frac{\partial \omega^2}{\partial (kR_0)} =  i\frac{a}{R_0} - \frac{ \gamma}{\rho R_0^3} \left[ (kR_0)  - 2(kR_0)^3  \right] \\
\rightarrow ~~ && i\frac{a}{R_0}\frac{\rho R_0^3}{ \gamma} 
 =   (kR_0)-2(kR_0)^3   \,. \label{eq:maxima}
 \eeqs 
This cubic equation for $kR_0$ produces three solutions for the most unstable non-dimensional wavenumber, $(kR_0)^*$. Two out of three solutions are complex conjugate pairs. When we evaluate the absolute value of the imaginary part of the three solutions, these two conjugate solutions are identical. Therefore, Fig. \ref{most}(a) shows only two lines: one solution in the red line and two conjugate pairs in the blue line. 
In the limit of large $kR_0\rightarrow \infty$, the first term on the right-hand side of Eq. (\ref{eq:maxima}) can be negligible compared to the second term. Therefore, we can find the most unstable wavenumber $(kR_0)^*$ as
\beq 
(kR_0)^* = (-i)^{1/3} \left( \frac{\rho R_0^3}{ 2\gamma} \frac{a}{R_0} \right)^{1/3} = \frac{i}{2}  {\mathrm{Bo}^{1/3}} \,,
\eeq 
where the Bond number is defined as $\mathrm{Bo} = {\rho a (2R_0)^2}/{\gamma}$. It is worth noting that the typical Bond number is defined with a gravitational constant, $g$, instead of $a$. However, we redefine the Bond number with the object's acceleration, $a$, to characterize the effect of acceleration against capillarity. The above relation shows that the most unstable nondimensional wavenumber $(kR_0)^*$ scales as the 1/3 power of the Bond number or the 1/3 power of the acceleration. This 1/3 power of acceleration is confirmed in the limit of high acceleration as shown in Fig. \ref{most}(a).  

Finally, the most unstable growth rate, $\omega^*$, can be obtained by plugging $kR_0 = (kR_0)^* = (i/2) \, \mathrm{Bo}^{1/3}$ in Eq. (\ref{pinoff-disp}) as
\beqs 
\omega^{*2} &=& -\frac{a}{R_0} \frac{\mathrm{Bo}^{1/3}}{2} + \frac{ \gamma}{2\rho R_0^3} \left[ \frac{\mathrm{Bo}^{2/3}}{4} + \frac{\mathrm{Bo}^{4/3}}{16}  \right] \nonumber  \\ 
&=&  \left[ -\frac{a}{R_0} + \frac{ \gamma}{2\rho R_0^3}  \left( \frac{\rho a R_0^2}{ 2\gamma} + \left( \frac{\rho a R_0^2}{ 2\gamma} \right)^{1/3} \right) \right] \left( \frac{\rho a R_0^2}{ 2\gamma} \right)^{1/3} \nonumber \\
&=&  \left[ -\frac{3}{4} \frac{a}{R_0} + \frac{ \gamma}{2\rho R_0^3}   \left( \frac{\rho a R_0^2}{ 2\gamma} \right)^{1/3}  \right] \left( \frac{\rho a R_0^2}{ 2\gamma} \right)^{1/3} \,.
\eeqs 
In the limit of high Bond numbers (i.e., $\mathrm{Bo} \gg 1$), the most unstable growth rate scales as
\beq 
\omega^{*2} \sim a^{4/3} ~\rightarrow~~ \mathrm{Im}[\omega^*] \sim a^{2/3} \,.
\eeq 
This 2/3 power of acceleration is confirmed in Fig. \ref{most}(b). 

To calculate the theoretical pinch-off time, we consider both the most unstable wavenumber and growth rate. The imaginary components of both wavenumber and frequency will contribute to changing the magnitude of variables. In some literature, the imaginary part of the wavenumber is called as a spatial growth rate and the imaginary part of the frequency as a temporal growth rate \cite{Keller1973,driessen2014control}. Likewise, both imaginary parts will describe spatial and temporal growth rates in the column radius or other variables. 
For the column radius, $R = R_0 - C \exp(ikz - i\omega t) = R_0 - C \exp(-\mathrm{Im}[(k)^*z]  + \mathrm{Im}[\omega^*] t) \exp(i \mathrm{Re}[(k)^*z] - i \mathrm{Re}[\omega^*] t)$. At a characteristic distance $z = 2R_0$, the characteristic timescale is proportional to $\mathrm{Im}[2(kR_0)^*]/\mathrm{Im}[\omega^*]$. 
Then, we assume the theoretical pinch-off time as 
\beq 
t_\mathrm{pinch-off} \simeq \frac{ \mathrm{Im} [2 (kR_0)^*] }{\mathrm{Im} [\omega^*] } = \left( {\frac{16}{3 \sqrt[3]{2}} } \right)^{1/2}  \left( \frac{\rho R_0^5}{\gamma} \right)^{1/6} a^{-1/3} \label{t_pinch} \,.
\eeq 
The nondimensional pinch-off time is given as 
\beq 
\tilde{t}_\mathrm{pinch-off} =  \frac{2}{\sqrt{3} }  \, \mathrm{Bo}^{-1/3} \label{Nt_pinch} \,.
\eeq 
Here, the pinch-off time is normalized by the capillary time defined as $\sqrt{\rho (2R_0)^3/\gamma}$\,. 
As a remark, our theoretical pinch-off time is different from the previous prediction ($\tilde{t}_\mathrm{pinch-off} = \sqrt{8\pi/9} \, \mathrm{Bo}^{-1/2}$) in \cite{Kim2018}.

\subsection{Comparison with experiments}
In this section, we will validate our theoretical pinch-off time with physical experiments and biological measurements. Two data sets are from physical experiments performed using ethanol or water by pulling up either a cylinder \cite{Gart2015} or a sphere \cite{Kim2018}. Biological data are from animal lapping behaviors; one cat \cite{Reis2010} and 19 dogs \cite{Gart2015}.  

\begin{figure}
\centering
\includegraphics[width=.9\textwidth]{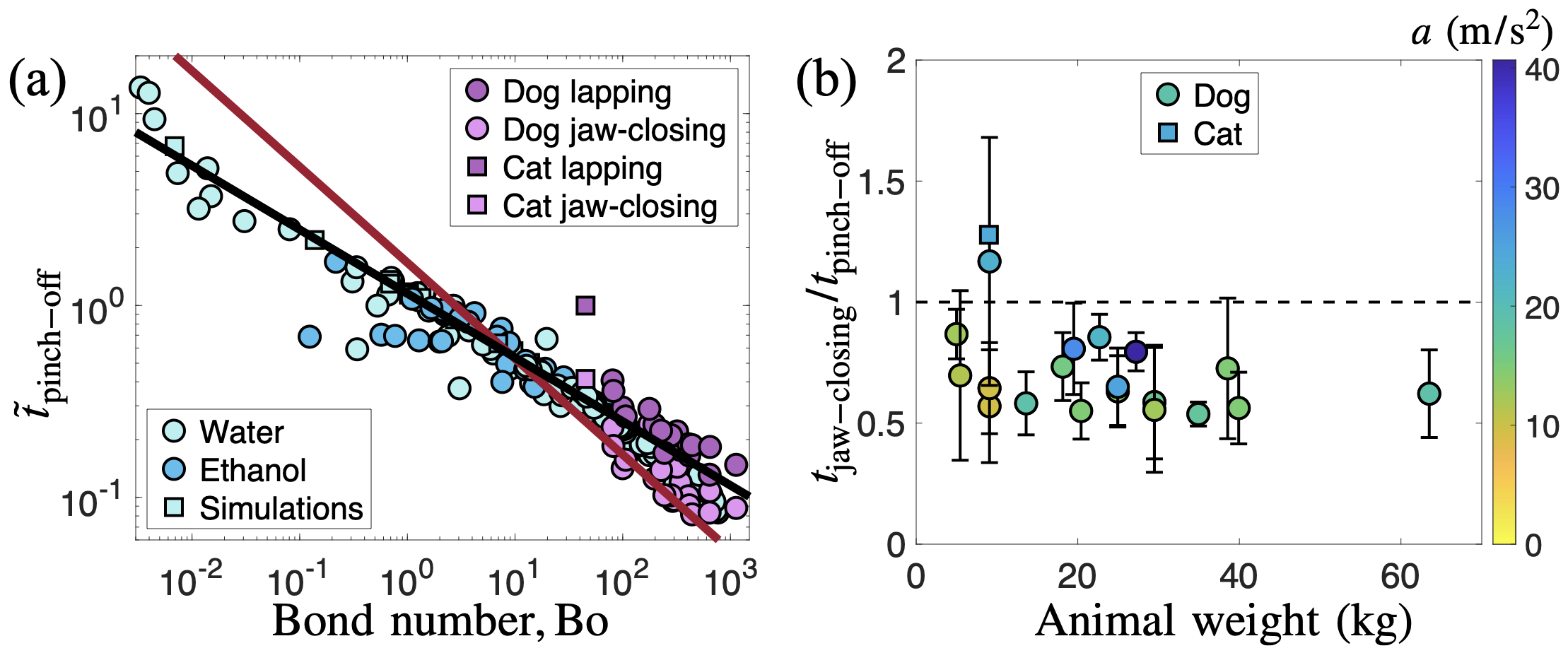}
\caption{ (a) Dimensionless pinch-off time ($\tilde{t}_\mathrm{pinch-off}$) versus the acceleration-based Bond number ($\mathrm{Bo} =\rho a (2R_0)^2/\gamma$). 
A black line is from our present theory, $\tilde{t}_\mathrm{pinch-off}={2/\sqrt{3}} \, \mathrm{Bo}^{-1/3}$, whereas a purple line indicates the previous theory,  $\tilde{t}_\mathrm{pinch-off}=\sqrt{8\pi/9} \, \mathrm{Bo}^{-1/2}$. It shows that our new theoretical pinch-off time works quite well with both physical experiments and dog \& cat lapping. 
(b) Jaw-closing timescale normalized by our predicted pinch-off time versus animal weight. As shown here, most animals close their jaws before the pinch-off time ($t_\mathrm{jaw-closing} < t_\mathrm{pinch-off}$). 
}
\label{predict}
\end{figure}

Figure \ref{predict}(a) shows the nondimensional pinch-off time vs. the Bond number. This nondimensional pinch-off time is measured as the experimental pinch-off time divided by the capillary time. Blue circle and square symbols indicate experimental \cite{Kim2018,Gart2015} and numerical \cite{Kim2018} studies, respectively. Light blue symbols are from experiments with water and dark blue ones are from experiments with ethanol. For dogs and a cat, there are two different pinch-off timescales measured; jaw-closing time and lapping time. The jaw-closing time is defined as a time difference between when the tongue exits from the water surface and when the animal closes its jaw. The lapping time is measured as a half of the inverse of the lapping frequency. These two types of time were measured from recorded videos of dog or cat lapping's. In the previous study \cite{Gart2015}, dogs lap fresh water; however, we have not measured the fluid properties on site, but assume that fresh water has $\rho = 1000$ kg/m$^3$ and $\gamma = 0.072$ N/m. In the study of cats \cite{Reis2010}, a cat laps milk mixed with tuna juice a little. We also have not measured the fluid properties either, but assume that milk has $\rho = 1030$ kg/m$^3$ and $\gamma = 0.052$ N/m based on reported milk properties \cite{Whitnah1959,fox2015physical}. Additionally, we do not have many cat data due to the lack of video footage from other angles to measure both the tongue radius and kinematics. In Fig. \ref{predict}(a), light purple symbols are the jaw-closing time, while dark purple symbols are the lapping time. As shown here, our new theory of the -1/3 power (black line) works quite well with experimental data compared to the previous theory of the -1/2 power (red line). For biological data, they follow the -1/3 power of the Bond number quite well. However, the cat data deviate a bit from the trend. We will explain the possible error for the cat in the next paragraph.  

Figure \ref{predict}(b) shows the jaw-closing time normalized by the theoretical pinch-off time versus animal weight. The theoretical pinch-off time in Eq. (\ref{t_pinch}) is calculated based on the acceleration and radius of the tongue measured in dog and cat experiments. We find that most normalized jaw-closing times ($t_\mathrm{jaw-closing}/t_\mathrm{pinch-off}$) are less than 1, which indicates that animals close their jaws just before the column pinch-off. However, there are two data points above 1: one is the 9-kg dog and the other is the 9-kg cat. There is no special feature on the 9-kg dog, but this dog exhibits a wide range of jaw-closing times (see the large error bar in the plot). For the cat, the jaw-closing time is also larger than the theoretical pinch-off time. There are two complications in cat drinking. As reported in the previous study \cite{Reis2010}, the cat's tongue accelerates first and then decelerates a bit near the end of the lapping period. Hence, the acceleration of cat drinking is not constant over time, which is quite different from dog drinking with constant acceleration. Our theory based on the constant acceleration might not be able to explain the cat lapping. In addition, we did not have video footage from multiple cameras to accurately measure the tongue kinematics and radius in contact with a liquid bath. 


\section{Conclusion}
In this present study, we showed how the theoretical pinch-off time changes with the acceleration of an object exiting a liquid bath. We performed the stability analysis of both 1D and 2D Euler equations for the liquid column, and ended up with a similar dispersion relation. Then, the pinch-off time was determined by the most unstable wavenumber and growth rate without a constant wavenumber assumption. The theoretical pinch-off time follows the -1/3 power of the Bond number, which is in good agreement with the experimental data. Moreover, this result possibly predicts the lapping frequency of dogs and cats, which indicates that animals modulate their lapping and jaw-closing times to catch the water column before the pinch-off to maximize the water intake. 

\section{Acknowledgement}
The author thanks Jake Socha and Sean Gart to allow me to use the photo and data of dog lapping, and SeongJin Kim for providing the previously published data. This research is partially supported by National Science Foundation (NSF) Grant CBET-2002714.

\bibliography{NSF21}


\end{document}